\documentclass[aps,prl,twocolumn,superscriptaddress,floatfix]{revtex4-1}
\usepackage{float}
\usepackage{amsfonts}
\usepackage{amsmath}
\usepackage{amssymb}
\usepackage{graphicx}
\usepackage{listings}
\usepackage{bm}
\usepackage{bbold}
\usepackage{braket}
\usepackage{todonotes}
\usepackage{hyperref}
\usepackage{epsfig}
\usepackage{times}

\begin{document}

\title{Anderson localization of pairs  in bichromatic optical lattices}
\author{Gabriel Dufour}
\affiliation{Laboratoire Mat\' eriaux et Ph\'enom\`enes Quantiques, Universit\' e Paris Diderot-Paris 7 and CNRS, UMR 7162, 75205 Paris Cedex 13, France}
\author{Giuliano Orso}
\affiliation{Laboratoire Mat\' eriaux et Ph\'enom\`enes Quantiques, Universit\' e Paris Diderot-Paris 7 and CNRS, UMR 7162, 75205 Paris Cedex 13, France}
\date{\today}

\begin{abstract}

We investigate the formation of  bound states made of two interacting atoms moving in a one dimensional (1D) quasi-periodic optical lattice. We  derive the quantum phase diagram for Anderson localization of 
both attractively and repulsively bound pairs.  We calculate the pair binding energy  and 
 show analytically that its behavior as a function of the interaction strength depends crucially on the nature -extended, multifractal, localized- of the single-particle atomic  states.
 Experimental implications of our results are discussed.
%
\end{abstract}

\maketitle

\textsl{Introduction.} The study of Anderson localization\cite{anderson58} (AL) of particles in  disordered or quasi-disordered potentials is a cornerstone in condensed matter physics. The recent observations \cite{Aspect2008, Inguscio2008,Demarco_AL,aspect3D} of AL in clouds of  ultra-cold atoms \cite{zwerger}  in the presence of  laser speckles  or quasi-periodic optical potentials has opened new prospects  \cite{reviewDis} to study  the rich interplay of disorder and interaction effects in a highly controlled way. Preliminary experimental results \cite{interaction_delocalization} have been obtained for weakly interacting 1D Bose gases subject to a quasi-periodic potential. Theoretical studies \cite{tommaso} have also addressed the emergence of a Bose-glass phase in these systems.
Interestingly, AL of light waves was recently observed \cite{lahini} in similar 1D configurations using photonic lattices and 
non-linear effects have been investigated.

In this Letter we address  the problem of two interacting particles moving in a quasi-periodic 1D optical lattice
\cite{doublebutterfly,katsanos,romer}.  We derive the phase diagram for Anderson localization of  
both attractively and repulsively bound pairs \cite{antibound} which can be promptly accessed in 
current experiments with ultra-cold atoms. 
We  calculate the pair  binding energy  $E_b$ and show analytically that when the constituent single particle states
become critical, $E_b$ exhibits an anomalous power law exponent [cfr. Eq.\ref{general}] as a function of the interaction strength. We also  introduce a  simple variational ansatz yielding very accurate results for the binding energy in all regimes.  
Since pairing  is found to dramatically modify the localization properties of atoms, our approach provides also an important tool to investigate
disordered Fermi superfluids  \cite{bec-bcs, demelo,tezuka}  undergoing the so called BCS-Bose Einstein Condensation 
(BEC) crossover \cite{zwerger}.




%

Let us start by writing the Schrodinger equation for the  single particle states in the absence of interaction:
\begin{equation}
\label{AApotential}
-J\phi(n+1)-J \phi(n-1)+V(n) \phi(n)=\varepsilon \phi(n),
\end{equation} 
where $J$ is the hopping rate between nearest sites, and $V(n)=\Delta   \cos(2\pi\beta n+\theta)$ is the quasi-periodic external potential. Here $\Delta$ is the potential strength, $\theta$ is a uniform (random) phase and  $\beta$ is the ratio between the wave-vectors of the two laser beams (see Ref.\cite{drese,michelemodugno2009} for details).  Eq.(\ref{AApotential}) is generally referred as the Harper \cite{harper} or the Aubry-Andr\' e model \cite{AubryAndre}.  For $\Delta=2J$ the $(\varepsilon,\beta)$ spectral diagram  is the well-known Hofstadter butterfly \cite{hof}  and for  irrational $\beta$  the system undergoes an AL transition.
In particular, for $\Delta<2J$ all states are extended whereas for $\Delta>2J$ all states are exponentially localized   with a localization length $\xi=1/\ln(\Delta/2J)$ independent of the energy.
For $\Delta=2J$  all eigenstates are critical and exhibit multi-fractal behavior \cite{multifractality}. 
More explicitly, for large but finite system sizes $L$, one can associate to each wave-function an infinite set of  fractal dimensions $D_q$ which are defined from the scaling behavior of  $\sum_{n=1}^L |\phi(n)|^{2 q} \propto L^{-D_q(q-1)}$. In particular the correlation dimension $D_2$ controls the behavior of the inverse participation ratio near the transition point.
 The associated energy spectrum is also characterized by a set of fractal exponents, which can be introduced from the dependence of the bandwidth of a given level on the system size, $\varepsilon^a-\varepsilon^p\propto L^{-\gamma} $, where $\varepsilon^a,\varepsilon^p$ are the eigenenergies calculated with, respectively, periodic and antiperiodic boundary conditions.
For example, for  $\beta=(\sqrt{5}-1)/2$ (golden ratio), the single particle ground state has  $D_2=0.329$ \cite{FSS} and $\gamma=2.374$ \cite{multifractality}.


In the presence of on-site (Hubbard) interactions $\hat U=U\sum_{m}|m,m\rangle \langle m,m|$ between the two particles,
the Schr\" odinger equation can be written as 
$(E-\hat H_0)|\psi\rangle=\hat U|\psi\rangle$, where $E$ is the energy
and $\hat H_0$ is the non interacting two-body Hamiltonian.  By applying the  Green's function operator $\hat G_E=(E-\hat H_0)^{-1}$ to both sides of the equation, we find 
  $|\psi\rangle=\hat G_E \hat U|\psi\rangle$. 
Projecting over the state $|n,n^\prime\rangle$  gives 
\begin{equation}\label{formalism1}
\psi(n,n^\prime)=U\sum_m \langle n,n^\prime|\hat G_E|m,m\rangle \psi(m,m),  
\end{equation}
where $\psi(n,n^\prime)=\langle n,n^\prime|\psi\rangle$ is the amplitude of the two-particle wave-function. 
The matrix elements in  Eq.(\ref{formalism1}) can be obtained by expressing 
the Green's function operator  in terms of the non interacting eigenbasis, 
$\hat G_E=\sum_{r,s} (E-\varepsilon_r-\varepsilon_s)^{-1}|\phi_r,\phi_s\rangle\langle \phi_r,\phi_s|$, 
where $\phi_r$ are the eigenstates of Eq.(\ref{AApotential}) with energy $\varepsilon_r$
written in ascending order $\epsilon_1<\epsilon_2<...<\epsilon_L$; this gives
\begin{equation}\label{kernel}
 \langle n,n^\prime|\hat G_E|m,m\rangle=\sum_{r,s}  \frac{\phi_r(n) \phi_s(n^\prime) \phi_r^*(m) 
\phi_s^*(m)}{E - \varepsilon_r - \varepsilon_s}.
\end{equation}

Eq.(\ref{formalism1}) and (\ref{kernel}) show that for contact interactions the  two-body wave-function 
  can be entirely 
reconstructed from the diagonal terms $f(m)=\psi(m,m)$. By setting $n^\prime=n$ in Eq.(\ref{formalism1}),
we end up with the  following eigenvalue problem \cite{PRAOrso} 
 \begin{equation}\label{integral}
 \frac{1}{U}f(n)=\sum_{m} K_E(n,m) f(m ),
 \end{equation} 
where the kernel is $K_E(n,m)= \langle n,n|\hat G_E|m,m\rangle$ and the  eigenvalue is $\lambda=1/U$.  
For values of the energy below the two-particle  non interacting spectrum $(E<2\epsilon_1)$, the eigenvalues 
$\lambda$ are all negative corresponding to \textsl{attractively bound} states, since $U<0$ and the pair has a finite size.  In particular the ground state energy  is obtained by varying $E$ until the lowest eigenvalue $\lambda$
equals $1/U$.
When the energy is above the two-particle  non interacting spectrum $(E>2\epsilon_L)$, the eigenvalues $\lambda$ are all positive corresponding to \textsl{repulsively bound} states,  since $U>0$. 
In particular, a wave-function $f(n)$ describing an attractively bound state with energy $E$ represents also a repulsively bound state with energy $-E$ provided the uniform phase $\theta$ is shifted by $\pi$.

For numerical stability, the irrational number $\beta$ is usually  expressed as the limit of a continued fraction. For definiteness we take $\beta=(\sqrt{5}-1)/2$  which can be approximated as 
$\beta\simeq  F_{j-1}/F_j$, where $F_j$ are Fibonacci numbers (defined by $F_0 = 0, F_1 = 1$ and $ F_j = F_{j-1} + F_{j-2} $ for $j \geq 2$), and  $j$ is  sufficiently large.
 Finite size effects are minimized by 
fixing the length of the chain to  $L=F_j$ and imposing \textsl{periodic boundary conditions}. 
Hereafter we set $J=1$ and measure all energies in units of the 
tunneling rate.


\textsl{Anderson localization of pairs}.  
Eq.(\ref{integral}) should be interpreted as an \textsl{effective} single-particle Schr\" odinger equation for the center-of-mass motion of the pair. In the absence of the  quasi-periodic potential $(\Delta=0)$,  the solutions of  Eq.(\ref{integral}) are delocalized Bloch states $f(n)=e^{i k n}/\sqrt{L}$, since  
the quasi-momentum $k$ of the pair is conserved.
 In the opposite atomic limit $(\Delta\rightarrow \infty)$, the kernel in Eq.(\ref{integral}) becomes diagonal in real space, $K_E(n,m)=\delta_{n,m}/(E-2V(n))$, implying that the pair is localized $f(n)=\delta_ {n,n_0}$,  since the two particles must  be in the same site  $n_0$ to bind together. 
 
\begin{figure}[tb]
\includegraphics[width=0.85\columnwidth]{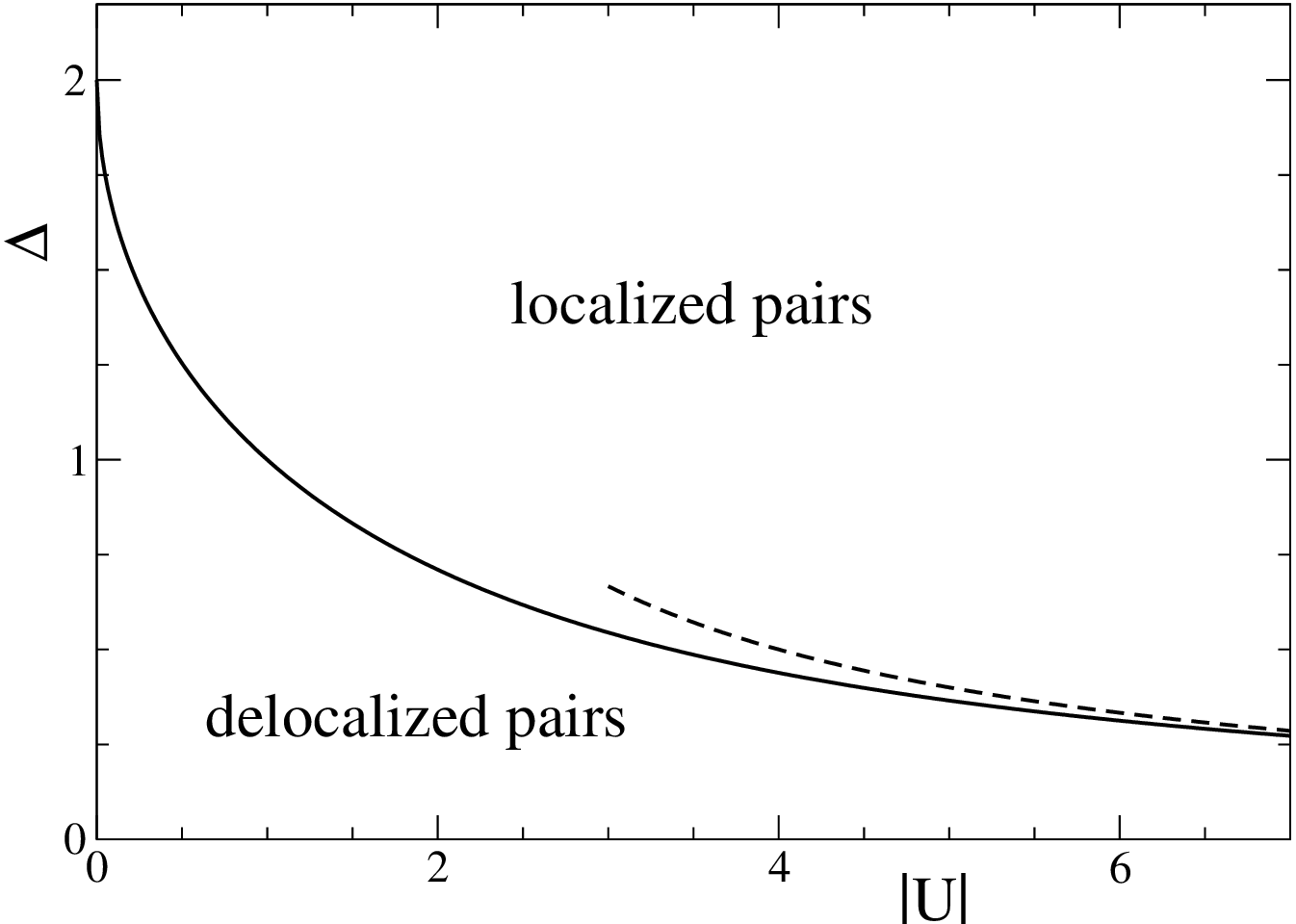} 
\caption{ Quantum phase diagram for Anderson localization of pairs in a 1D quasi-periodic lattice. The critical potential strength  is plotted as a function of the interaction  (energy units $J=1)$.    The dashed line corresponds to the asymptotic regime of tightly bound pairs where AL occurs at $\Delta_\textrm{cr}=-2/U$. The phase diagram applies to both attractively and repulsively bound pairs.}
\label{fig_anderson}
\end{figure}

To calculate the critical strength $\Delta=\Delta_\textrm{cr}$  where AL occurs,  
we introduce the  inverse participation ratio 
$\alpha_\textrm{p}=\sum_{n=1}^{L}|f(n)|^4$ of the pair. In the localized phase  $\alpha_\textrm{p}$ is always finite whereas in the extended phase  it vanishes  in the thermodynamic limit as $\alpha_\textrm{p}\propto L^{-1}$.  
For $U<0$, we calculate $\alpha_p$ for the ground state solution of Eq.(\ref{integral}) 
with  fixed energy $E$ and increasing values of the potential strength.
We then identify  $\Delta_\textrm{cr}$  as the inflection point of the obtained curve $\alpha_\textrm{p}(\Delta)$ \cite{noteIPR}. Repeating the procedure for different  $E$  gives the quantum phase diagram shown in Fig.\ref{fig_anderson}.
For vanishing interactions the pair breaks and $\Delta_\textrm{cr}\rightarrow 2$ as expected for single atoms. As the interaction strength 
increases (in modulus), we see that the localized phase extends progressively to smaller and smaller values of $\Delta$. 
This result can be better understood in the strong coupling regime $|U|\gg \Delta,1$ where pairs are tightly bound
objects with large effective mass. By expanding the kernel in powers of $E^{-1}$ up to the  third order included, we find
\begin{eqnarray}
K_E(n,m)&\simeq &\delta_{n,m} \left(\frac{1}{E}+\frac{2V(n)}{E^2}+\frac{4 V(n)^2+4}{E^3}\right)\nonumber\\
&&+\frac{2}{E^3}\delta_{n,m\pm1}, \label{kernelSC}
\end{eqnarray}
showing that to a first approximation $E\sim U$ [see Eq.(\ref{integral})].
For tightly bound pairs we expect $\Delta_\textrm{cr} \ll 1$, so we can safely neglect the $V^2$ term in the right hand side of Eq.(\ref{kernelSC}). As a consequence, the motion of the pair is governed by an equation like (\ref{AApotential}), but
with  an effective  tunneling rate $J_\textrm{eff}=-2/E\ll 1$ and a potential strength $\Delta_\textrm{eff}=2\Delta$.
Since AL occurs at $\Delta_\textrm{eff}=2 J_\textrm{eff}$, we obtain  $\Delta_\textrm{cr}=-2/E \simeq -2/U$, as shown in Fig.(\ref{fig_anderson}) with the dashed line.
We also notice that  the phase boundary between localized and delocalized molecular states does not depend on the choice of the angle $\theta$. Consequently, the phase diagram in Fig.(\ref{fig_anderson}) \textsl{applies unchanged} also to repulsively bound pairs for $U>0$. 

\begin{figure}[tb]
\includegraphics[width=0.85\columnwidth]{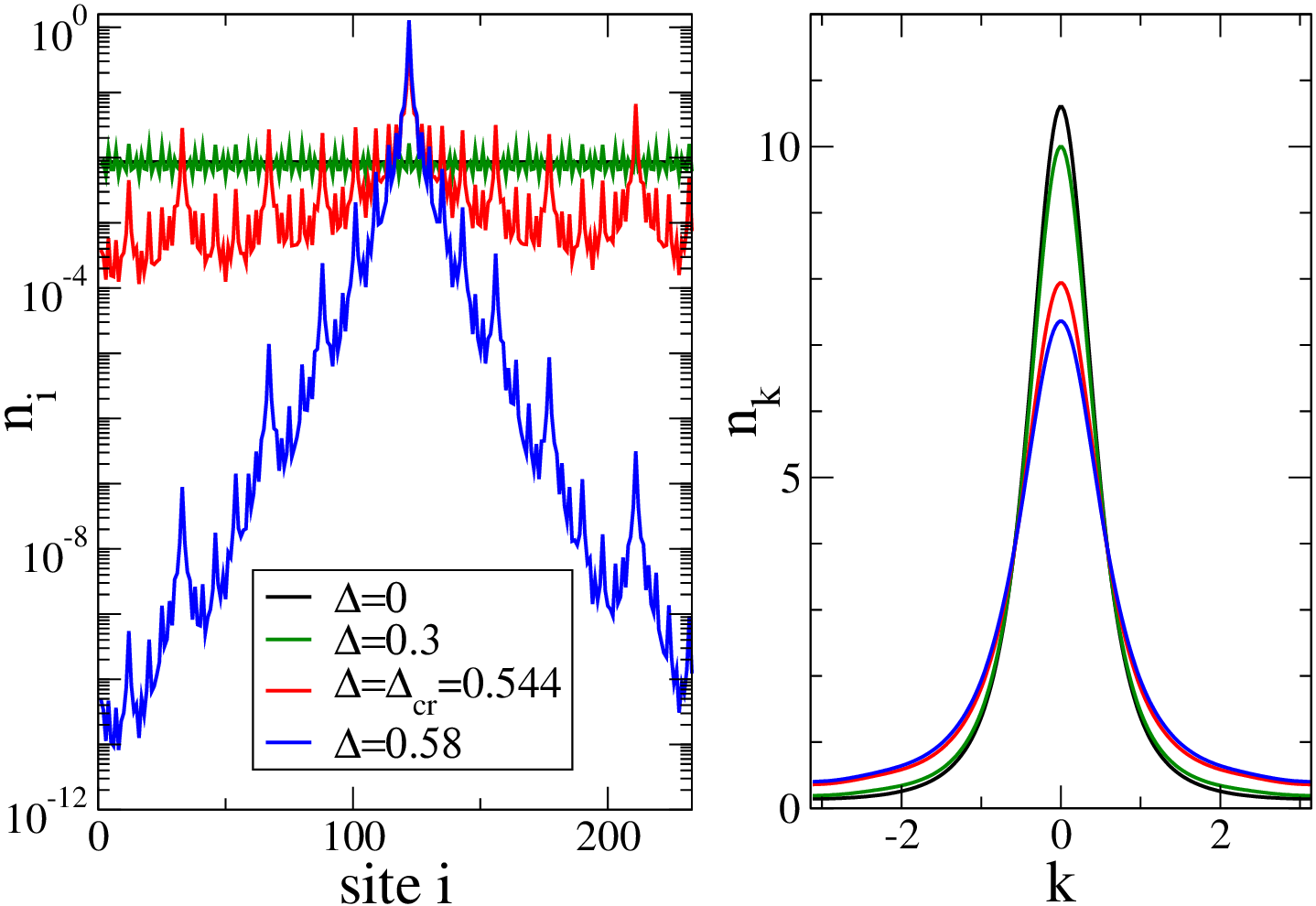} 
\caption{(Color online)  Crossing the phase boundary of Fig.\ref{fig_anderson}. Local density $n_i$ (left panel) and quasi-momentum distribution $n_k$ [normalized to $\int n_k dk/(2\pi)=2$] (right panel) of atoms inside molecules for fixed value of $U=-3$ and increasing values of the potential strength (starting from the top). At $\Delta=\Delta_\textrm{cr}= 0.544$ the molecular state becomes Anderson localized. 
We used $\theta=\pi/5$ and $L=233$ \cite{noteFig2} . }
\label{fig_local}
\end{figure}

Since in a bound state the averaged distance between the two atoms is finite,  
the density profile of atoms  must change from extended 
  to exponentially localized as the phase boundary in Fig.(\ref{fig_anderson}) is crossed.  
  To see this, from Eqs (\ref{integral}) and (\ref{formalism1}) we reconstruct the ground state wave-function, normalized to $\sum_{i,j}|\psi(i,j)|^2=1$. 
In  Fig.(\ref{fig_local}) we plot  the calculated local density  $n_i=2 \sum_m |\psi(i,m)|^2$ (left panel) and the quasi-momentum distribution $n_k=2 \sum_{i,j,m} \psi(i,m)^* \psi(j,m) e^{i k (j-i)}$  (right panel) of the two constituent atoms  for fixed $U=-3$ and for increasing values of $\Delta$. We emphasize that  the observed localization of the density profile is \textsl{induced by interactions}, as non interacting atoms would  remain delocalized for any $\Delta<2$.
In contrast the quasi-momentum distribution of atoms behaves smoothly across the phase boundary, as shown in 
the right panel of Fig.(\ref{fig_local}).

%
%
%
%

\begin{figure}[tb]
\includegraphics[width=0.85\columnwidth]{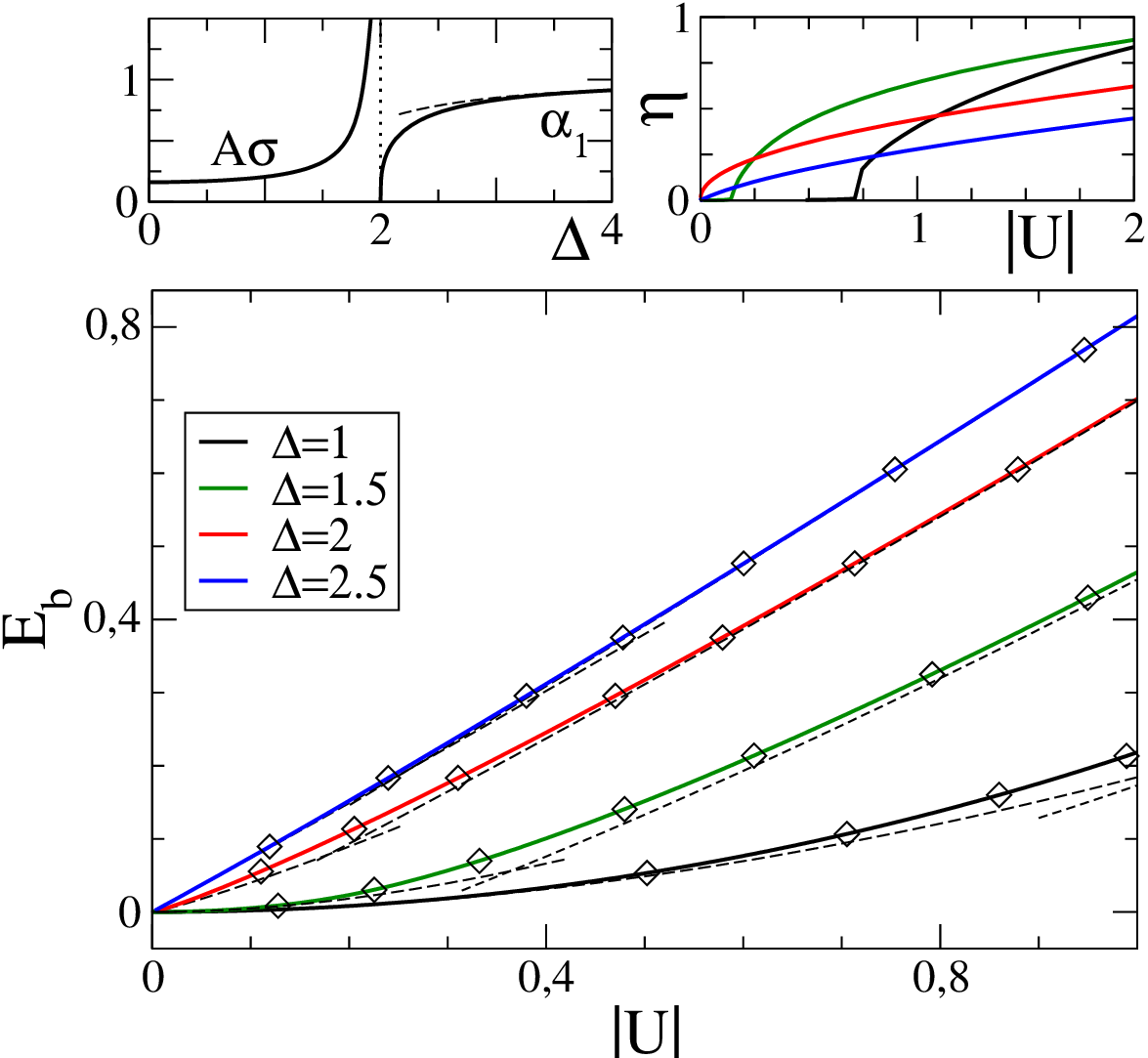} 
\caption{ (color online)  Main panel: Binding energy of molecules versus interaction  for increasing values
of the  potential strength $\Delta=1\;(\textrm{bottom}),1.5,2,2.5$. 
The asymptotic behaviors for strong  and weak interactions, obtained from  Eq.(\ref{approx})] and Eq.(\ref{weakU}) respectively, are shown with dashed lines. Diamonds correspond to the predictions of the variational ansatz (\ref{varansatz}). 
Upper left panel:  $A\sigma$ [cfr. Eq.(\ref{asymp_weakUINT})] and $\alpha_1$ [cfr.  Eq.(\ref{atomic})], as a function of the 
potential strength. 
Also shown is the asymptotic behavior of $\alpha_1$ for $\Delta\gg 1$.
Right panel: best variational parameter $\eta$ as a function of the interaction strength $|U|$ and for the same values of $\Delta$ as in the main panel ($\Delta=2$ is the top curve near $U=0$). }
\label{fig_binding} 
\end{figure}

\textsl{Binding energy.} The pair binding energy $E_b$  is defined in the usual way from $E=2 \varepsilon_1-E_b$, where $\varepsilon_1$ is the ground state energy for a single particle [see Eq.([\ref{AApotential})]. In Fig.\ref{fig_binding} (main panel) we show  our numerical results for the
binding energy as a function of the attractive interaction $U$ and for increasing values of 
$\Delta$ (solid lines). 
We see that in general the quasi-periodic potential favors the formation of molecules, because interaction effects are enhanced. For 
$|U|\Delta \gg 1$, the molecule is trapped near the minimum of the external potential 
and we can treat the hopping term perturbatively. To  second order included we find  $E=U-2\Delta+4/(U-\Delta(1-\cos(2\pi \beta)))$, from which we get
\begin{equation}\label{approx}
E_b=-U+2\Delta + 2 \varepsilon_1-\frac{4}{U-\Delta(1-\cos(2\pi \beta))}.
\end{equation}

For weak interactions $|U|\ll  1$,  the binding energy can be calculated analytically starting from the \textsl{free-particle} ansatz  $f(n)=\phi_1(n)^2$ in Eqs.  (\ref{kernel}) and (\ref{integral}). 
Taking into account that
$\sum_n \phi_r(n) \phi_s(n) =\delta_{r,s}$, we find
\begin{equation}
\label{weakU}
1\simeq -U\sum_r \frac{\alpha_{r}}{E_b+2\varepsilon_r-2\varepsilon_1},
\end{equation}
where $\alpha_r=\sum_n \phi_r(n)^2 \phi_1(n)^2$ are overlap functions. In particular $\alpha_1=\sum_n\phi_1(n)^4$ is the  inverse participation ratio  of the single particle ground state.

 For $\Delta<2$ all single-particles states are delocalized and the overlaps vanish in the thermodynamic limit  as $\alpha_r\propto L^{-1}$. For small $E_b$, the leading contribution in Eq.(\ref{weakU}) comes from the
low-lying energy states, where the  overlaps become uniform, $\alpha_r L \simeq \sigma$, 
$\sigma$ being a constant.
  Going to continuous variables $\sum_r \rightarrow L \int \rho(\epsilon)d\epsilon$ , 
where $\rho(\epsilon)$ is the single particle density of states, we can write Eq.(\ref{weakU}) as 
$1=-U\sigma  \int_0^\infty  \rho(\epsilon)/(E_b+2 \epsilon)$,
with $\epsilon=\varepsilon_r-\varepsilon_1$.
Taking into account that
at low energy  $\rho(\epsilon)\sim A/\sqrt{\epsilon}$, where $A$ is a constant that depends smoothly on $\Delta$, we find that 
\begin{equation}\label{asymp_weakUINT}
E_b\simeq \frac{\pi^2}{2}A^2 \sigma^2 U^2,
\end{equation}
 showing that the binding energy depends quadratically on the interaction strength  for molecules built from delocalized states.
 The asymptotic behavior (\ref{asymp_weakUINT}) is shown with dashed lines in Fig.\ref{fig_binding}. We see that by increasing $\Delta$ its range of validity shrinks to weaker
 and weaker interactions. At the critical point, 
  the coefficient $\sigma$ in Eq.(\ref{asymp_weakUINT})    diverges as $\sigma\propto (2-\Delta)^{-(1-D_2)}$ as shown in the upper left panel of Fig.\ref{fig_binding}. 
  
On the other hand for $\Delta >2$, the single particle  spectrum is point-like and 
all the states are  localized.  Taking into account that the low-lying states have vanishing overlaps 
$\alpha_r\simeq \alpha_1 \delta_{r,1}$, from Eq.(\ref{weakU}) we find
the linear in $U$ dependence
\begin{equation}\label{atomic}
E_b\simeq |U| \alpha_1,
\end{equation}
typical of systems with discrete energy levels.  In the atomic limit $\Delta \gg 1$, where all states are localized within few lattice sites, the hopping term in Eq.(\ref{AApotential}) can be treated perturbatively, yielding $\alpha_1\simeq 1-4/(\Delta (1-\cos (2 \pi \beta)))^2$. For weaker $\Delta$, the inverse participation ratio decreases and vanishes at the critical point as $\alpha_1\propto (\Delta-2)^{D_2}$ , as shown in the upper left panel of Fig.\ref{fig_binding}.

For $\Delta=2$,   where multi-fractality of the single particle states emerges,  a fit to our numerical data reveals  a power law behavior 
$E_b=C|U|^\delta$, with $C=0.720$ and an  anomalous exponent $\delta=1.161$.
 In particular we find numerically that the overlap functions and the energy differences of the single particle ground state scale with the system size as,  respectively, $\alpha_r\propto L^{-D_2}$ and  $\epsilon_r -\epsilon_1\propto L^{-\gamma}$, with
$D_2$ and $\gamma$ defined above. By substituting $\alpha_r=U_r L^{-D_2}$
 and $\epsilon_r -\epsilon_1=V_r L^{-\gamma}$  in Eq.(\ref{weakU}), where $U_r,V_r$ are independent of the system size (provided $r\ll L$), we find  
\begin{equation}\label{multi}
\frac{1}{U} =-L^{\gamma-D_2}g(E_b L^\gamma),
\end{equation}
where $g(x)=\sum_{r=1}^\infty U_r/(x+2 V_r)$. Since the left hand side of Eq.(\ref{multi}) is $L-$independent, $g(x)$
 must be a power law $g(x)\propto x^{-1/\delta}$ from which we obtain   %
\begin{equation}\label{general}
\delta=\frac{\gamma }{\gamma-D_2},
\end{equation}
which is fully consistent with our numerics. Eq.(\ref{general}) is   actually valid for any values of $\Delta$.
Indeed, for  $\Delta<2$ it yields $\delta=2$ because
in the single particle metallic phase $D_2=1$ and $\gamma=2$, whereas for  $\Delta>2$ one finds $\delta=1$  since $D_2=0$ in the  localized phase, in agreement with Eqs.(\ref{asymp_weakUINT}) and (\ref{atomic}), respectively. 
 Eq.(\ref{weakU})   nevertheless predicts a wrong numerical prefactor $C_\textrm{ansatz}=0.58$ at the critical point. This comes from the fact that 
 for any finite interaction the molecule is localized, as shown in the phase diagram of Fig.(\ref{fig_anderson}),  whereas our naive free-particle ansatz remains critical. 

\textsl{Variational ansatz}. 
For finite $U$, we consider the following generalized ansatz
\begin{equation}\label{varansatz}
f_\textrm{var}^\eta(n)=\phi_1^2(n) e^{-\eta d(n,n_0)},
\end{equation}
where $\eta>0$ is a variational parameter, $n_0$ corresponds to a site where the quasi-periodic 
potential takes its minimum value and $d(n,n_0)$ represents the distance between  the two  sites. Within this class of states, the lowest eigenvalue of Eq.(\ref{integral}) is given by
\begin{equation}
\frac{1}{U}=\underset{\eta}{\textrm{min}} \frac{\sum_{n,m} f_\textrm{var}^\eta(n)K_E(n,m) f_\textrm{var}^\eta(m)}{\sum_n
|f_\textrm{var}^\eta(n)|^2}.
\end{equation}
The obtained results are shown with diamonds in the main panel of Fig.\ref{fig_binding} and are in  very good agreement with the full numerical data for all explored values of $U$ and $\Delta$. In particular  we
recover the correct numerical prefactor $C_\textrm{var}=0.720$ at $\Delta=2$.
In the upper right panel of Fig.\ref{fig_binding} we  see that $\eta$ becomes finite before the phase boundary in Fig.\ref{fig_anderson} is reached.  
To understand this point notice that the ansatz (\ref{varansatz}) applies only \textsl{close} to $n_0$ so that
$\eta$ should not be confused with the inverse localization length of the molecule, which instead concerns  the \textsl{tails} of the wave-function.

\textsl{Implications for experiments.}
The phase diagram (\ref{fig_anderson}) can be explored experimentally with sufficiently dilute bosonic  
clouds of atoms by adding a second incommensurate lattice to the original set-up of Ref.\cite{antibound}.
 Notice that the density profiles in Fig.\ref{fig_local} are calculated at equilibrium whereas  in Ref.\cite{Inguscio2008} the spatial distribution of atoms  is measured after  letting the particles expand along the 1D bichromatic lattice. 
In this case we  expect that  atoms  bound into pairs will stop expanding for $\Delta>\Delta_\textrm{cr}$ \cite{tezukaBIS}. 
Finally we point out that the pair binding energy can also be measured experimentally
in optical lattices  by  rf spectroscopy \cite{moritz,mit}.

\textsl{Acknowledgements}. We thank M. Valiente, G. Modugno and Y. Lahini for useful comments on the manuscript.



\begin{thebibliography}{}
\bibitem{noteIPR} For  $L=233$  the uncertainty in the value of $\Delta_\textrm{cr}$  is  below 0.2$\%$.
\bibitem{tezukaBIS} This is indeed seen in recent numerical simulations  of  fermionic systems \cite{tezuka}
although the values of $\Delta_\textrm{cr}$ found there are  smaller (up to $15\%$)  probably due many-body effects.
\bibitem{noteFig2} The choice of  $\theta$ determines the position of the localization peak in the chain but does not
affect the other observables presented in this paper, since $L$ is sufficiently large.
\end{thebibliography}
\end{document}